\documentclass{acm_sen_article}

\usepackage{color}
\begin{document}
\title{Automated Verification and Synthesis of \\ Embedded Systems using Machine Learning}

\author{Lucas Cordeiro \\
%Department of Computer Science\\
Department of Computer Science, University of Oxford, UK\\
E-mail: lucas.cordeiro@cs.ox.ac.uk
%\and
%Eddie B. de Lima Filho \\
%Science, Technology, and Innovation Center\\ for the Industrial Pole of Manaus, Brazil \\
%Email: eddie@ctpim.org.br}
%\IEEEauthorblockN{Homer Simpson}
%\IEEEauthorblockA{Twentieth Century Fox\\
%\and
%\IEEEauthorblockN{James Kirk\\ and Montgomery Scott}
%\IEEEauthorblockA{Starfleet Academy\\
%San Francisco, California 96678--2391\\
%Fax: (888) 555--1212}
}


%Telephone: (800) 555--1212, Fax: (888) 555--1212}

\maketitle

\begin{abstract}
The dependency on the correct functioning of embedded systems is rapidly growing, mainly due to their wide range of applications, such as micro-grids, automotive device control, health care, surveillance, mobile devices, and consumer electronics. Their structures are becoming more and more complex and now require multi-core processors with scalable shared memory, in order to meet increasing computational power demands. As a consequence, reliability of embedded (distributed) software becomes a key issue during system development, which must be carefully addressed and assured. 
The present research discusses challenges, problems, and recent advances to ensure correctness and timeliness regarding embedded systems. Reliability issues, in the development of micro-grids and cyber-physical systems, are then considered, as a prominent verification and synthesis application. In particular, machine learning techniques emerge as one of the main approaches to learn reliable implementations of embedded software for achieving a \textit{correct-by-construction} design.
\end{abstract}

%\IEEEpeerreviewmaketitle

%--------------------------------------
\section{Introduction}
%--------------------------------------

Generally, embedded computer systems perform dedicated functions with a high degree of reliability. They are used in a variety of sophisticated applications, which range from entertainment software, such as games and graphics animation, to safety-critical systems, including nuclear reactors and automotive controllers~\cite{Kopetz11}. Embedded systems are ubiquitous in modern day information systems, and are also becoming increasingly important in our society, especially in micro-grids, where reliability and carbon emission reduction are of paramount importance~\cite{xu15}, and in cyber-physical systems (CPS), which demand short development cycles and again a high-level of reliability~\cite{leeCPS2}. As a consequence, human life has also become more and more dependent on the services provided by this type of system and, in particular, their success is strictly related to both service relevance and quality. 

\begin{figure}[!t]
	\centering
	\includegraphics[scale=0.34]{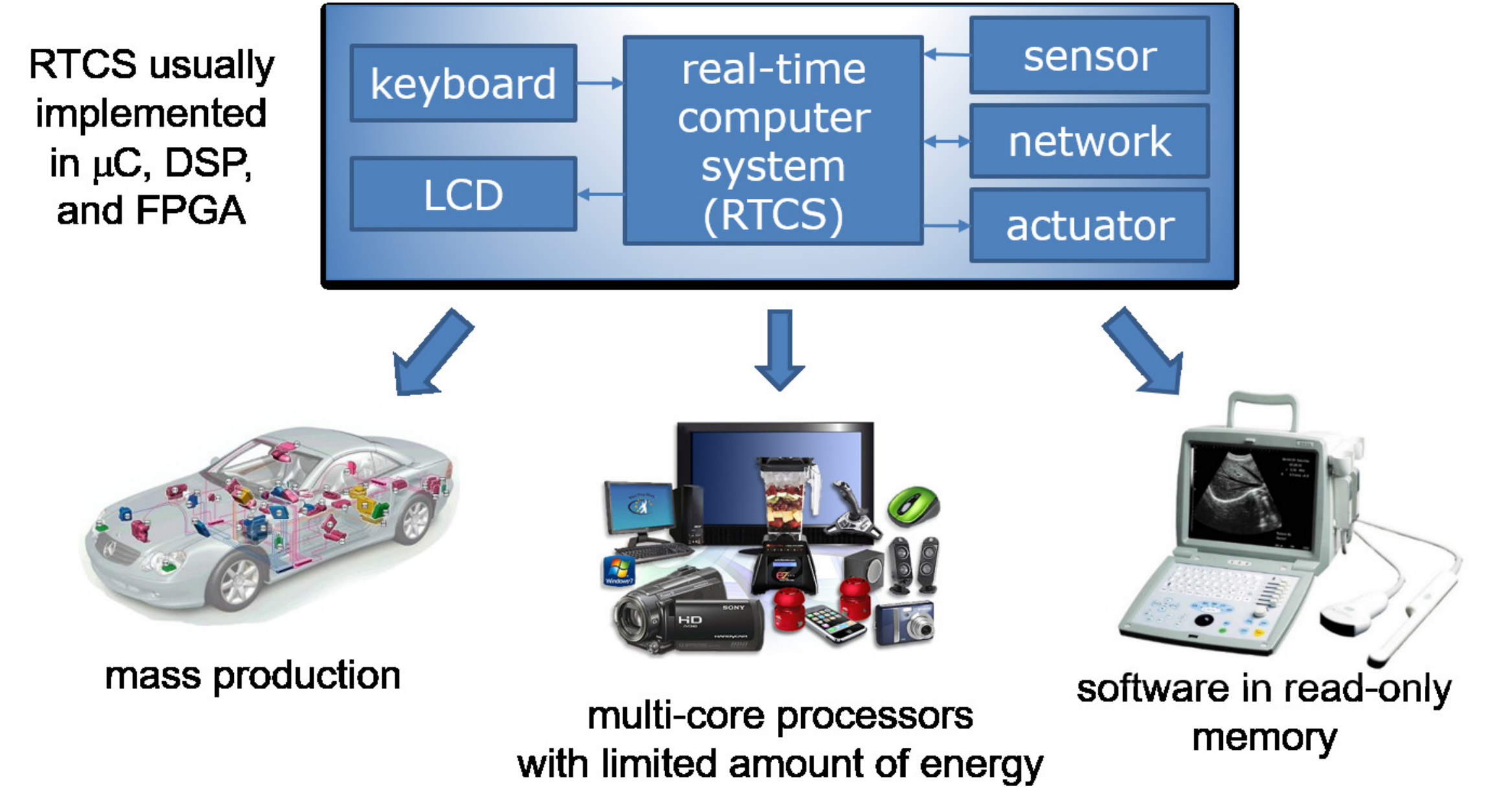}
	\caption{An embedded system is part of a well-specified larger system (intelligent product). \vspace{-2ex}}
	\label{intelligent-product}
\end{figure}

Figure~\ref{intelligent-product} shows embedded systems examples, which typically consist of a human-machine interface ({\it e.g.}, keyboard and LCD), a processing unit ({\it e.g.}, real-time computer system), and an instrumentation interface ({\it e.g.}, sensor, network, and actuator) \cite{Kopetz11}. Indeed, many current embedded systems, such as unmanned aerial vehicles (UAVs)~\cite{groza2015formal} and medical monitoring systems~\cite{Cordeiro09}, become interesting solutions only if they can reliably perform their target tasks. Besides, when physical interaction with the real world is needed, which happens in CPS, additional care must be taken, mainly when human action is directly replaced, as in vehicle driving. Regarding the latter, even human-in-the-loop feedback control can be employed, which raises deeper concerns w.r.t. reliability of human behavior modeling and system implementation.
%For instance, UAVs are a trend on military missions, due to the absence of pilots; however, an incorrect plan execution may cost civilian lives, which is unacceptable. In addition, wrong disease diagnosis or condition-evaluation reports have the potential to compromise patients' health, with serious consequences.
%

Consequently, it is important to go beyond design correctness and also address behavior correctness, which may be performed by incorporating system models. In particular, these models can be used for synthesizing a given system, ensuring that all needed functions are correctly implemented and the correct behavior exhibited, {i.e.}, the system is indeed correct by its method of construction~\cite{Abate17}. Here, machine learning emerges as a powerful technique to automatically learn the correct behavior of the system, which must provably satisfy a given correctness specification $\sigma$. Specifically, synthesizers can use $\sigma$ as starting point and then incrementally produce a sequence of candidate solutions that satisfy $\sigma$, by integrating deductive methods with inductive inference (learning from counterexamples)~\cite{Seshia15}. As a result, a given candidate solution can be iteratively refined to match the specification $\sigma$ based on a counterexample-guided learning approach.

%--------------------------------------
\section{Verification and Synthesis \\ Challenges for Embedded Systems}
\label{Verification-Challenges} 
%--------------------------------------

\begin{figure}[h]
	\centering
	\includegraphics[scale=0.35]{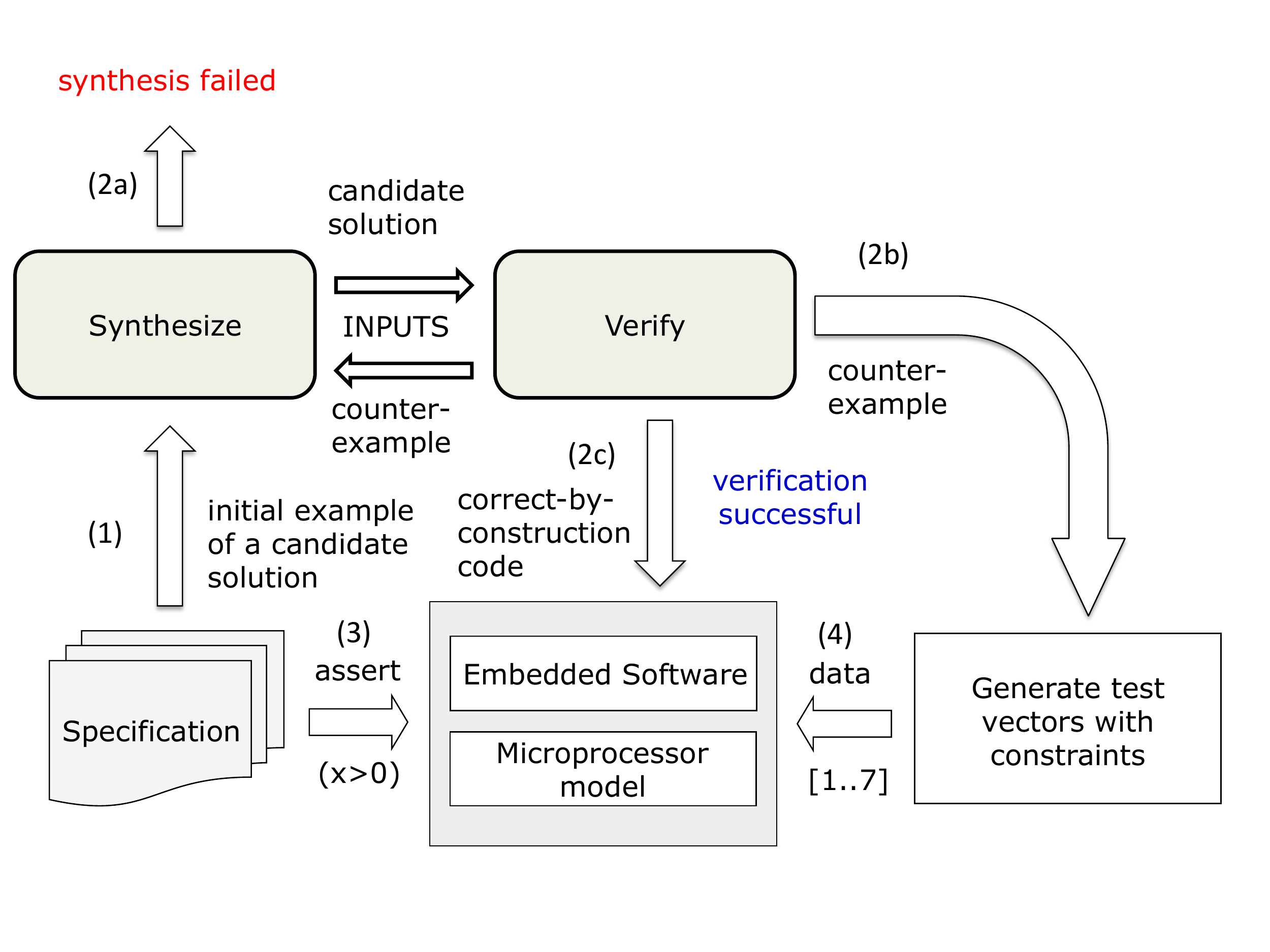}
	\caption{Verification and synthesis methodologies for embedded systems. \vspace{-3ex}}
	\label{verification-methodologies}
\end{figure}

State-of-the-art verification methodologies for embedded systems generate test vectors (with constraints) and use assertion-based verification and high-level processor models, during simulation~\cite{Behrend15}, as shown in Figure~\ref{verification-methodologies}. 
Here, the main challenges regarding the verification of embedded systems lie on improving coverage, pruning the state-space exploration during verification, and incorporating system models, which allow specific checks regarding system behavior and not only code correctness. Additionally, embedded system verification raises additional challenges, such as: (1) time and energy constraints; (2) handling of concurrent software; (3) platform restrictions; (4) legacy designs; (5) support to different programming languages and interfaces; and (6) handling of non-linear and non-convex optimization functions.

Indeed, the first two aspects are of extreme relevance in micro-grids and cyber-physical systems, in order to ensure reliability, which is a key issue for (smart) cities, industries, and consumers, and the third one is essential in systems that implement device models, such as digital filters and controllers, which present a behavior that is highly dependent on signal inputs and outputs and whose deployment may be heavily affected by hardware restrictions. The fourth aspect is inherent to a large number of embedded systems from  telecommunications, control systems, and medical devices. In particular, software developed for those systems has been extensively tested and verified, and also optimized for efficiency over years. Therefore, when a new product is derived from a given platform, a lot of legacy code is usually reused for reducing development time and improving code quality. The fifth aspect is related to the evolution of development processes and technologies, which may delay the application of suitable verification and synthesis approaches if verifiers and synthesizers do not support different programming languages and interfaces. The last one is related to the widespread use of embedded systems in autonomous vehicle navigation systems, which demand optimization solving during their execution for a wide range of functions, including non-linear and non-convex optimization functions.

Those challenges place difficulties for developing (reliable) synthesizers for embedded systems, especially for CPS and micro-grids, where the controlled object ({\it e.g.}, physical plant) typically exhibits continuous behavior whereas the controller (usually implemented by a real-time computer system) operates in discrete time and over a quantized domain (cf. intelligent product in Figure~\ref{intelligent-product}). In particular, synthesizers for those systems need to consider the effects of the quantizers (A/D and D/A converters), when a digital equivalent of the controlled object is considered, {\it i.e.}, a model of their physical environment. Additionally,  finite-precision arithmetic and their related rounding errors need to be considered when correct-by-construction code is generated for embedded systems. The main challenge lies on exploiting effectively and efficiently counterexamples provided by verifiers to automatically learn reliable embedded software implementations (cf. Figure~\ref{verification-methodologies}).

%--------------------------------------
\section{Research Problem (RP)}
\label{Research-Problem}
%--------------------------------------

This research statement tackles six major problems in computer-aided verification and synthesis for embedded systems, which are (partially) open in current published research.

\textbf{(RP1)} provide suitable encoding into SMT, which may extend the background theories typically supported by SMT solvers, with the goal of reasoning accurately and effectively about realistic embedded (control) software.

\textbf{(RP2)} exploit SMT techniques to leverage bounded model checking of multi-threaded software, in order to mitigate the state-explosion problem due to thread interleaving.
	
\textbf{(RP3)} prove correctness and timeliness of embedded systems, by taking into account stringent constraints imposed by hardware.
	
\textbf{(RP4)} incorporate knowledge about system purpose and associated features to detect system-level and behavior failures.

\textbf{(RP5)} provide tools and approaches capable of addressing different programming languages and application interfaces, with the goal of reducing the time needed to adapt current verification techniques to new developments and technologies.

\textbf{(RP6)} develop automated synthesis approaches that are algorithmically and numerically sound, in order to handle embedded (control) software that is tightly coupled with the physical environment by considering uncertain models and FWL effects.

%--------------------------------------
\section{Current Achievements and \\ Future Trends}
\label{achievements}
%--------------------------------------

In order to support SMT encoding \textbf{(RP1)}, Cordeiro, Fischer, and Marques-Silva proposed the first SMT-based BMC for full C programs, called Efficient SMT-Based Context-Bounded Model Checker (ESBMC)~\cite{Cordeiro12}, which was later extended to support C++, CUDA, and Qt-based consumer electronics applications. This approach was also able to find undiscovered bugs related to arithmetic overflow, buffer overflow, and invalid pointer, in standard benchmarks, which were later confirmed by the benchmarks' creators ({\it e.g.}, NOKIA, NEC, NXP, and VERISEC). Other SMT-based BMC approaches have also been proposed and implemented in the literature~\cite{Armando06,MerzFS12}, but the coverage and verification time of all existing ones are still limited to specific classes of programs, especially for those that contain intensive floating-point arithmetic and dynamic memory allocation. One possible research direction is to bridge the gap between BMC tools and SMT solvers to propose background theories and develop more efficient decision procedures to handle specific classes of programs.

The SMT-based BMC approach proposed by Cordeiro, Fischer, and Marques-Silva was further developed to verify correct lock acquisition ordering and the absence of deadlocks, data races, and atomicity violations in multi-threaded software based on POSIX and CUDA libraries \textbf{(RP2)}~\cite{CordeiroF11,Pereira15}, considering monotonic partial-order reduction and state-hashing techniques, in order to prune the state-space exploration. Recent advances for verifying multi-threaded C programs have been proposed to speed up the verification time, which significantly prune the state-space exploration; however, the class of concurrent programs ({\it e.g.}, OpenCL and MPI) that can be verified is still very limited. One possible research direction is to further extend BMC of multi-threaded programs via Sequentialization~\cite{Inverso14} and also analyze interpolants to prove non-interference of context switches~\cite{McMillan11}.

Novel approaches to model check embedded software using \textit{k}-induction and invariants were proposed and evaluated in the literature \textbf{(RP3)}, which demonstrate its effectiveness in some real-life embedded-system applications~\cite{Gadelha15}. However, the main challenge still remains open, {\it i.e.}, to compute and strengthen loop invariants to prove program correctness and timeliness in a more efficient and effective way, in order to be competitive with other model-checking approaches~\cite{Rocha17}. In particular, invariant-generation algorithms have substantially evolved over the last years, with the goal of discovering inductive invariants of programs or continuously refine them during verification~\cite{Beyer15}. Yet there is still a lack of studies for exploiting the combination of different invariant-generation algorithms ({\it e.g.}, interval analysis, linear inequalities, polynomial equalities and inequalities) and how to strengthen them.% during verification. %, in order to ensure system robustness w.r.t. implementation aspects.

State-of-the-art SMT-based context-BMC approaches were extended to verify overflow, limit cycle, stability, and minimum phase, in digital systems \textbf{(RP4)}. Indeed, digital filters and controllers were tackled, in order to verify system-level properties of those systems, specified as linear-time temporal logic~\cite{JMorse15,Bessa17}. In particular, a specific UAV application was tackled, with the goal to verify its attitude controllers. In general, however, there is still a lack of studies to verify system-level properties related to embedded systems; emphasis should be given to micro-grids and cyber-physical systems, which require high-dependability requirements for computation, control, and communication. Additionally, the application of automated fault detection, localization, and correction techniques to digital systems represents an important research direction to make BMC tools useful for engineers.

Although ESBMC was extended to support C/C++ and some variants \textbf{(RP5)}, new application interfaces and programming languages are often developed, which require suitable verifiers. Indeed, it would be interesting if a new programming language model could be loaded, which along with a BMC core could check different programs. Some work towards that was already presented by~\cite{Sousa17}, which employed operational models for checking Qt-based programs from consumer electronics. In summary, the BMC core is not changed, but instead an operational model, which implements the behavior and features of Qt libraries, is used for providing the new code structure to be checked. Such research problem is closely related to the first one \textbf{(RP1)} and has the potential to devise a new paradigm in software verification.

State-of-the-art synthesis approaches \textbf{(RP6)} for embedded (control) systems typically disregard the platform in which the embedded system software operates and restrict itself to generate code that do not take into account FWL effects. However, the synthesized system must include the physical plant to avoid serious system's malfunctioning (or even a catastrophe) due to the embedded (control) software, {\it e.g.}, the Mars Polar Lander did not account for leg compressions prior to landing~\cite{Jackson16}. Research in this direction has made some progress to design, implement, and evaluate an automated approach for generating correct-by-construction digital controllers that is based on state-of-the-art inductive synthesis techniques~\cite{Abate17}. However, there is still little evidence whether that approach can scale for larger systems modeled by other types of representations ({\it e.g.}, state-space). Another research direction for synthesizers is to automatically produce UAV trajectory and mission planning code, by taking into account system's dynamics and nonholonomic constraints. As a result, verifiers and synthesizers need to handle a wide range of functions, including non-linear and non-convex optimization problems~\cite{Araujo17}. Machine learning techniques could be employed here to learn from counterexamples, {\it i.e.}, in the inductive step, synthesizers could learn the model from raw data, and in the deductive step, the model could be applied to predict the behaviour of new data~\cite{Alur13}.

%--------------------------------------
\section{Conclusions}
\label{conclusions}
%--------------------------------------

This research statement presented the main challenges related to the verification of design correctness, in embedded systems, and also raised some important side considerations about synthesis. 
%
%In particular, it emphasizes that stringent constraints imposed by the underlying hardware ({\it e.g.}, real-time, memory allocation, interrupts, and concurrency), along with system behavior models, must be considered during verification and synthesis. 
%Additionally, there is little evidence that model checking embedded software using \textit{k}-induction (and invariants), which extends BMC-based approaches from falsification to verification, can be applied to formally verify correctness and timeliness of embedded systems. 
%
Given that software complexity has significantly increased in embedded products, there is still the need for stressing and exhaustively covering the entire system state space, in order to verify low-level properties that have to meet the application's deadline, access memory regions, handle concurrency, and control hardware registers. Besides, there is a trend towards incorporating knowledge about the system to be verified, which may take software verification and synthesis one step further, where not only code correctness will be addressed, but also full system reliability. Finally, it seems interesting to provide behavioral models when new application interfaces or programming language features are used, in order to extend the capabilities of current verification tools, without changing the core BMC module.

As future perspective, the main goal of this research is to extend BMC as a verification and synthesis tool for achieving correct-by-construction embedded system implementations. Special attention will be given to CPS and modern micro-grids, considering small-scale versions of a distributed system, so that reliability and other system-level properties ({\it e.g.}, carbon emission reduction in smart cities) are amenable to automated verification and synthesis, probably through behavior models.


%\section*{Acknowledgment}

%The authors thank M. Dangl for reviewing a draft version of this paper.~This research was supported by CNPq $475647$/$2013$-$0$ grant.

\end{document}